\RequirePackage{lineno}  

\documentclass[prl,aps,floatfix,showpacs,superscriptaddress,twocolumn]{revtex4}

\usepackage{graphicx,dcolumn,url}
\usepackage{lineno}
\def \Et {{E}_{T}}
\def \met  {\,/\!\!\!\!E_{T}}

\begin{document} 
                                            
\title{
Search for a Neutral Higgs Boson Decaying to a $W$ Boson Pair in $p\bar{p}$
Collisions at $\sqrt{s} = 1.96$~TeV }
\affiliation{Institute of Physics, Academia Sinica, Taipei, Taiwan 11529, Republic of China} 
\affiliation{Argonne National Laboratory, Argonne, Illinois 60439} 
\affiliation{Institut de Fisica d'Altes Energies, Universitat Autonoma de Barcelona, E-08193, Bellaterra (Barcelona), Spain} 
\affiliation{Baylor University, Waco, Texas  76798} 
\affiliation{Istituto Nazionale di Fisica Nucleare, University of Bologna, I-40127 Bologna, Italy} 
\affiliation{Brandeis University, Waltham, Massachusetts 02254} 
\affiliation{University of California, Davis, Davis, California  95616} 
\affiliation{University of California, Los Angeles, Los Angeles, California  90024} 
\affiliation{University of California, San Diego, La Jolla, California  92093} 
\affiliation{University of California, Santa Barbara, Santa Barbara, California 93106} 
\affiliation{Instituto de Fisica de Cantabria, CSIC-University of Cantabria, 39005 Santander, Spain} 
\affiliation{Carnegie Mellon University, Pittsburgh, PA  15213} 
\affiliation{Enrico Fermi Institute, University of Chicago, Chicago, Illinois 60637} 
\affiliation{Joint Institute for Nuclear Research, RU-141980 Dubna, Russia} 
\affiliation{Duke University, Durham, North Carolina  27708} 
\affiliation{Fermi National Accelerator Laboratory, Batavia, Illinois 60510} 
\affiliation{University of Florida, Gainesville, Florida  32611} 
\affiliation{Laboratori Nazionali di Frascati, Istituto Nazionale di Fisica Nucleare, I-00044 Frascati, Italy} 
\affiliation{University of Geneva, CH-1211 Geneva 4, Switzerland} 
\affiliation{Glasgow University, Glasgow G12 8QQ, United Kingdom} 
\affiliation{Harvard University, Cambridge, Massachusetts 02138} 
\affiliation{Division of High Energy Physics, Department of Physics, University of Helsinki and Helsinki Institute of Physics, FIN-00014, Helsinki, Finland} 
\affiliation{University of Illinois, Urbana, Illinois 61801} 
\affiliation{The Johns Hopkins University, Baltimore, Maryland 21218} 
\affiliation{Institut f\"{u}r Experimentelle Kernphysik, Universit\"{a}t Karlsruhe, 76128 Karlsruhe, Germany} 
\affiliation{High Energy Accelerator Research Organization (KEK), Tsukuba, Ibaraki 305, Japan} 
\affiliation{Center for High Energy Physics: Kyungpook National University, Taegu 702-701, Korea; Seoul National University, Seoul 151-742, Korea; and SungKyunKwan University, Suwon 440-746, Korea} 
\affiliation{Ernest Orlando Lawrence Berkeley National Laboratory, Berkeley, California 94720} 
\affiliation{University of Liverpool, Liverpool L69 7ZE, United Kingdom} 
\affiliation{University College London, London WC1E 6BT, United Kingdom} 
\affiliation{Centro de Investigaciones Energeticas Medioambientales y Tecnologicas, E-28040 Madrid, Spain} 
\affiliation{Massachusetts Institute of Technology, Cambridge, Massachusetts  02139} 
\affiliation{Institute of Particle Physics: McGill University, Montr\'{e}al, Canada H3A~2T8; and University of Toronto, Toronto, Canada M5S~1A7} 
\affiliation{University of Michigan, Ann Arbor, Michigan 48109} 
\affiliation{Michigan State University, East Lansing, Michigan  48824} 
\affiliation{Institution for Theoretical and Experimental Physics, ITEP, Moscow 117259, Russia} 
\affiliation{University of New Mexico, Albuquerque, New Mexico 87131} 
\affiliation{Northwestern University, Evanston, Illinois  60208} 
\affiliation{The Ohio State University, Columbus, Ohio  43210} 
\affiliation{Okayama University, Okayama 700-8530, Japan} 
\affiliation{Osaka City University, Osaka 588, Japan} 
\affiliation{University of Oxford, Oxford OX1 3RH, United Kingdom} 
\affiliation{University of Padova, Istituto Nazionale di Fisica Nucleare, Sezione di Padova-Trento, I-35131 Padova, Italy} 
\affiliation{LPNHE, Universite Pierre et Marie Curie/IN2P3-CNRS, UMR7585, Paris, F-75252 France} 
\affiliation{University of Pennsylvania, Philadelphia, Pennsylvania 19104} 
\affiliation{Istituto Nazionale di Fisica Nucleare Pisa, Universities of Pisa, Siena and Scuola Normale Superiore, I-56127 Pisa, Italy} 
\affiliation{University of Pittsburgh, Pittsburgh, Pennsylvania 15260} 
\affiliation{Purdue University, West Lafayette, Indiana 47907} 
\affiliation{University of Rochester, Rochester, New York 14627} 
\affiliation{The Rockefeller University, New York, New York 10021} 
\affiliation{Istituto Nazionale di Fisica Nucleare, Sezione di Roma 1, University of Rome ``La Sapienza," I-00185 Roma, Italy} 
\affiliation{Rutgers University, Piscataway, New Jersey 08855} 
\affiliation{Texas A\&M University, College Station, Texas 77843} 
\affiliation{Istituto Nazionale di Fisica Nucleare, University of Trieste/\ Udine, Italy} 
\affiliation{University of Tsukuba, Tsukuba, Ibaraki 305, Japan} 
\affiliation{Tufts University, Medford, Massachusetts 02155} 
\affiliation{Waseda University, Tokyo 169, Japan} 
\affiliation{Wayne State University, Detroit, Michigan  48201} 
\affiliation{University of Wisconsin, Madison, Wisconsin 53706} 
\affiliation{Yale University, New Haven, Connecticut 06520} 
\author{A.~Abulencia}
\affiliation{University of Illinois, Urbana, Illinois 61801}
\author{D.~Acosta}
\affiliation{University of Florida, Gainesville, Florida  32611}
\author{J.~Adelman}
\affiliation{Enrico Fermi Institute, University of Chicago, Chicago, Illinois 60637}
\author{T.~Affolder}
\affiliation{University of California, Santa Barbara, Santa Barbara, California 93106}
\author{T.~Akimoto}
\affiliation{University of Tsukuba, Tsukuba, Ibaraki 305, Japan}
\author{M.G.~Albrow}
\affiliation{Fermi National Accelerator Laboratory, Batavia, Illinois 60510}
\author{D.~Ambrose}
\affiliation{Fermi National Accelerator Laboratory, Batavia, Illinois 60510}
\author{S.~Amerio}
\affiliation{University of Padova, Istituto Nazionale di Fisica Nucleare, Sezione di Padova-Trento, I-35131 Padova, Italy}
\author{D.~Amidei}
\affiliation{University of Michigan, Ann Arbor, Michigan 48109}
\author{A.~Anastassov}
\affiliation{Rutgers University, Piscataway, New Jersey 08855}
% ...
\collaboration{CDF Collaboration}
\noaffiliation

\author{The CDF Collaboration}

\date{\today}

\begin{abstract}
We present the results of a search for standard model Higgs boson
production with decay to $WW^{*}$, identified through the leptonic
final states $e^{+}e^{-}\bar{\nu} \nu ,~e^{\pm}\mu^{\mp} \bar{\nu}
\nu$ and $\mu^{+}\mu^{-} \bar{\nu} \nu$. This search uses
360 $\rm{pb}^{-1}$ of data collected from $p\bar{p}$ collisions at
$\sqrt{s}$ = 1.96 TeV by the upgraded Collider Detector at Fermilab
(CDF II). We observe no signal excess and set 95\% confidence level
upper limits on the production cross section times branching ratio for
the Higgs boson to $WW^{*}$ or any new scalar particle with similar
decay products. These upper limits range from 5.5 to 3.2 $\rm{pb}$
for Higgs boson masses between 120 and 200 GeV/$c^2$.
\end{abstract}
                           

\maketitle


The Higgs mechanism is a leading candidate for electroweak symmetry
breaking and consequently for mass generation of the $W$ and $Z$
bosons without violation of local gauge invariance. A manifestation of
this mechanism is the existence of a neutral scalar particle, the
Higgs boson~\cite{higgs}, which has not been observed to date. Its
mass is a free parameter in the standard model (SM), but its couplings
to other particles of known mass are fully specified at tree
level. Direct searches at the CERN $e^{+}e^{-}$ collider (LEP) yielded
a lower limit for the Higgs boson mass of $m_{H}>$~114.4 GeV/$c^{2}$
at 95$\%$ confidence level (C.L.)~\cite{lep2lim}. Precision
electroweak measurements indirectly predict a Higgs boson mass of
$91^{+45}_{-32}$ GeV/$c^{2}$~\cite{ewprec}.

At the Tevatron, the dominant production mechanism for the SM Higgs
boson is gluon-gluon fusion through heavy quark loops. Branching
fractions for the various decay channels of the Higgs boson depend on
its mass. For masses below about $135\,{\rm GeV}/c^2$ the
dominant decay is $H \rightarrow b\bar{b}$, while heavier Higgs bosons
decay predominantly to $WW^\ast$~\cite{xsecnlo}, where $W^{*}$
indicates a $W$ boson that can be off mass-shell. For the $b \bar{b}$
decay mode, the requirement of associated production of the Higgs with
vector bosons ($p\bar{p}\rightarrow WH/ZH$) can greatly improve the
signal purity~\cite{run1_VH}. For the $WW^{*}$ decay mode, the
leptonic decays of $W$ bosons give a clean enough signature that the
inclusive single Higgs production process gives the best search
sensitivity. The next-to-leading order (NLO) production cross
section~\cite{xsecnlo} times branching ratio for a SM Higgs boson,
$\sigma(p\bar{p}\rightarrow H)\times$BR($H\rightarrow WW^{*})$, ranges
from 0.036 to 0.25 $\rm{pb}$ for Higgs masses of 110-200 GeV/$c^2$.
 
This Letter presents the results of a direct search for a Higgs boson in the
channel $gg \rightarrow H \rightarrow WW^{*}\rightarrow
\ell^{+}\nu\ell^{-}\bar{\nu}$ ($\ell=e, \mu, \tau$), identified by
the ``dilepton'' final states $e^{+}e^{-}, e^{\pm}\mu^{\mp}$ or
$\mu^{+}\mu^{-}$. We also include the efficiency for leptonically
decaying taus to e or $\mu$. This is the first search in this channel
by the CDF Collaboration. A similar search in this channel was
recently performed by the D\O\ Collaboration~\cite{hwwd0}. The data
sample used for this analysis was collected with the CDF II detector
at the Fermilab Tevatron between 2002 and 2004, and corresponds to an
integrated luminosity of approximately
360~$\rm{pb}^{-1}$~\cite{lumdetail}. For this integrated luminosity,
the cross section limits we are able to place on Higgs production are
a factor of approximately 10-50 larger than the SM expectation, based
on the NLO calculation.  However, the production cross-section can be
enhanced in extensions to the SM due to new particles e.g., a fourth
generation fermion family~\cite{4gen}, contributing at higher order to
the gluon-gluon fusion Higgs production process.


CDF II is a detector with approximate azimuthal and forward-backward
symmetry and it is fully described elsewhere~\cite{cdfdet}. It consists of a
charged-particle tracking system in a 1.4~T magnetic field and
segmented electromagnetic and hadronic calorimeters surrounded by muon
detectors. The electromagnetic and hadronic sampling calorimeters
surrounding the solenoid are used to measure the energy of interacting
particles in the pseudo-rapidity range $|\eta|<3.6$~\cite{eta}. The
calorimeters are divided into projective geometry towers. This
analysis uses both central ($|\eta|<1.1$) and end-plug detectors
($1.2<|\eta|<2.0$) to identify electron candidates. A set of drift
chambers located outside the central hadron calorimeters and another
set behind a 60 cm iron shield help detect muons in the region
$|\eta|<$ 0.6. Additional drift chambers and scintillation counters
detect muons in the region 0.6 $\leq|\eta|\leq 1.0$.

Events used for this analysis are collected using the following
triggers~\cite{trigele, trigplug}: an inclusive central electron
($|\eta|<1.1$) trigger requiring an electron with $E_{T}>$ 18
GeV, an inclusive central muon ($|\eta|<1.0$) trigger
requiring a muon with $p_{T}>$ 18 GeV/$c$, or a trigger for events
with a forward electron (1.2 $\leq|\eta|\leq 2.0$) with $E_{T}>$ 20
GeV and missing transverse energy, $~\met>$ 15 GeV~\cite{met}.


After the event reconstruction, event selection criteria which retain
high $H\rightarrow WW^{*}$ signal efficiency while minimizing the
effect of background contamination are applied. Some selection
requirements are mass dependent, as the event kinematics and topology
change as functions of $m_{H}$.

The selection requires two oppositely charged lepton candidates
consistent with originating from the same vertex, with $p_{T}>$ 20
GeV/$c$ for the trigger lepton and $p_{T}>$ 10 GeV/$c$ for the second
one. The leptons are also required to be isolated in both the
calorimeter and the tracking chamber~\cite{caliso}, and the dilepton
invariant mass $m_{\ell\ell}$ is required to be greater than
$16~{\rm{GeV}}/c^2$, in order to remove events from the
$c\bar{c}$/$b\bar{b}$ resonances.

After removal of events identified as cosmic rays or electrons from photon
conversions~\cite{trigele}, we count the jets~\cite{jetdef} with
$E_{T}>$ 15 GeV and $|\eta|<2.5$. Signal events do not typically have
high-$E_T$ jets in the final state, but can occasionally have
lower-$E_T$ jets from initial state gluon radiation. On the other
hand, $t\bar{t}$ pairs decay primarily to $W^{+}W^{-}b\bar{b}$ and
thus tend to have at least two jets in the final state. This
background is reduced by selecting only events satisfying one of the
following criteria: no jets with $E_{T}>15$ GeV, or only one jet with
$15<E_{T}<55$ GeV, or 2 jets, each with $15<E_{T}< 40$ GeV. Events
with more than two jets with $E_{T}>15$ GeV are also rejected.
 

After the selection criteria described above, the dominant surviving
background is Drell-Yan production of $\ell^{+}\ell^{-}$ pairs, which
is suppressed by requiring that $\met>m_{H}/4$. The events with
missing energy due to a mis-measurement of the jet energy, or
$Z\rightarrow\tau\tau$ events with missing energy arising from a
leptonic tau decay, are removed by requiring the azimuthal angle between
the $\met$ and the closest jet or lepton to be at least $20^\circ$, if
$\met<$ 50 GeV. To further reduce the large $Z/\gamma^{*}$ background,
the dilepton invariant mass is required to be $m_{\ell\ell} < m_{H}/2
- 5$ GeV/$c^2$. Finally, the scalar sum of the $p_{T}$ of the two
leptons and the $\met$ is required to be below the Higgs mass.

The kinematic cuts described above exploit the correlations in the $W$
pairs produced by the decay of a Higgs boson and suppress SM $WW$
production. These correlations are due to angular momentum
conservation in the decay of a spin-zero Higgs boson. Since $W$ bosons
decay into left-handed leptons and right-handed anti-leptons, and
since the W bosons in the decay $H\rightarrow WW^{*}$ have opposite
helicities, the final state lepton pairs and also the neutrino pairs
tend to be azimuthally aligned in Higgs decay. This implies that the
signal events tend to have smaller $m_{\ell\ell}$ and azimuthal angle
between leptons ($\Delta \phi$) and larger $\met$, as compared with
production of SM $WW$ pairs. These differences are further exploited
in the final stages of the analysis, when the $\Delta \phi$
distribution of the data is compared with the background and signal
predictions.


The acceptance for identifying $H\rightarrow WW^{*}\rightarrow
\ell\nu\ell\nu$ events with the above selection criteria is calculated
as a function of the Higgs boson mass using {\sc PYTHIA}~\cite{pythia}
Monte Carlo, after a {\sc GEANT}-based~\cite{geant} simulation of the
CDF detector response. The total acceptance is a product of the
geometric and kinematic acceptance, the lepton identification
efficiencies, the trigger efficiencies, and the topological cut
efficiencies. It does not include the branching fraction of $W$ leptonic
decays. The total acceptance ranges from 3.0$\%$ to 6.5$\%$, depending
on the Higgs mass, and is summarized in
Table~\ref{table:hwwacc}. Approximately 25\% of the expected signal
are $ee$ events, 25\% $\mu\mu$, and 50\% $e\mu$.

\begin{table}[h]
\caption{\label{table:hwwacc}
The branching ratio BR($H\rightarrow$ $WW^{*}$) and the total
acceptance of the signal after all the selection criteria. The total
acceptance is calculated with respect to the number of $p\bar{p}
\rightarrow H \rightarrow WW^{*}\rightarrow
\ell^{+}\nu\ell^{-}\bar{\nu}$  events. 
}

\begin{ruledtabular}
\begin{tabular}{lccccc}
$m_{H}$ (GeV/$c^2$) & 120 & 140& 160 & 180 &200 \\ \hline
BR($H\rightarrow$ $WW^{*})(\%)$ & 13 &  48 &  90 &  94 &  74  \\
Total acceptance($\%$) & 3.15& 4.56 & 6.47 & 6.41& 5.54\\

\end{tabular}
\end{ruledtabular}
\end{table}

The systematic uncertainty on the acceptance is 6$\%$ resulting from
uncertainties in the modeling of the initial state radiation by {\sc PYTHIA}
(3$\%$), and uncertainties on the gluon parton distribution functions
(4$\%$)~\cite{pdferror}, jet energy scale (1$\%$), track isolation
($<$2$\%$), electron and muon trigger efficiencies($<$1$\%$), and
electron and muon identification efficiencies (2$\%$).
In addition, a 6\% uncertainty on the integrated 
luminosity is applied to the expected number of events for all processes~\cite{lum}.


\begin{table*}[bth]
\caption{\label{table:grdtable}
 The expected number of signal and SM background events are
 presented. The number of events observed in the data, with the $m_H$
 dependent selection criteria, is also shown. The errors include all systematic effects.}

\begin{ruledtabular}
\begin{tabular}{lccccc}
$m_{H}$ (GeV/$c^{2}$) & 120 & 140& 160 & 180 &200 \\ \hline

$WW$           & 5.49 $\pm$ 0.66 &  7.98 $\pm$ 0.96 & 9.79 $\pm$ 1.18&  9.89 $\pm$ 1.19&  9.19 $\pm$ 1.11 \\
$Z/\gamma^{*}$ & 1.63 $\pm$ 0.42 & 1.01 $\pm$ 0.26 & 0.76 $\pm$ 0.20& 0.83 $\pm$ 0.21& 0.96 $\pm$ 0.25\\
$W+\rm{jets}$/$\gamma$  &4.57 $\pm$ 0.90 & 3.49 $\pm$ 0.81& 2.48 $\pm$ 0.69 & 1.70 $\pm$ 0.46& 1.20 $\pm$ 0.37\\
$WZ+ZZ$  & 0.25 $\pm$ 0.03 & 0.37 $\pm$ 0.05 & 0.40 $\pm$ 0.05 & 0.49 $\pm$ 0.07 & 1.16 $\pm$ 0.15 \\
$t\bar{t}$    & 0.12 $\pm$ 0.01  &  0.21 $\pm$ 0.02 & 0.35 $\pm$ 0.04 & 0.46 $\pm$ 0.05 & 0.58 $\pm$ 0.06 \\\hline
Total Background  &  12.06 $\pm$ 1.19&  13.08 $\pm$ 1.28&  13.78 $\pm$ 1.38&  13.37 $\pm$ 1.30&  13.09 $\pm$ 1.21 \\
\hline
$H\rightarrow$ $WW^{*}$ & 0.090 $\pm$ 0.008&  0.32 $\pm$ 0.03& 0.58 $\pm$ 0.05&  0.41 $\pm$ 0.03&  0.20 $\pm$ 0.02 \\\hline
Data  & 7 & 14 & 16 & 19 & 17\\
\end{tabular}
\end{ruledtabular}
\end{table*}

After all selection requirements, the background events come
predominantly from $WW$ pair production (about 70\% of the total for
$m_H = 160$ GeV/$c^2$)~\cite{WW}, $Z/\gamma^{*}$, $W + {\rm jets}$,
and $W+\gamma$. Smaller backgrounds include $WZ$, $ZZ$, and $t\bar{t}$
production. A summary of these contributions as a function of Higgs
mass is given in Table~\ref{table:grdtable}.  The diboson ($WW, WZ,
ZZ$), $Z/\gamma^{*}$  and $t\bar{t}$ backgrounds are determined using {\sc PYTHIA} Monte
Carlo, followed by the CDF II detector simulation. We normalize the
total number of events for these processes to recent theoretical
cross-sections~\cite{wwxsec, ttbarxsec}.
To estimate the $W+\gamma$ background we use a matrix element
generator~\cite{baur1} and use {\sc PYTHIA} for the initial state QCD
radiation and hadronization.

The background from $W + {\rm jets}$, where a jet or track is
misidentified as a lepton (electron or muon), is determined from the
data and called the ``fake background.'' We first determine the
probability that a jet with a large fraction of its energy deposited
in the electromagnetic calorimeter is misidentified as an electron,
and the probability that a minimum ionizing track is misidentified as
a muon. These probabilities are termed fake rates. The fake rate for
each lepton type is calculated using an average of four inclusive jet
samples (triggered with at least one jet with $E_{T} > $ 20, 50, 70 or
100 GeV). We subtract the contribution from sources of real leptons ($W$ and $Z$ decays)
and parametrize the fake rates as a function of jet transverse energy
(for electrons) or track transverse momentum (for muons). The
background is determined by weighting the jets from a data sample of
$(W \rightarrow \ell\nu) + \rm{jets}$ events by the fake rates.

For data events passing the previously described selection criteria,
we search for an excess of events with small azimuthal angle between
the leptons, $\Delta \phi$. A binned likelihood is used to compare the
azimuthal angle distribution in the data with a combination of
expected distributions from the SM background
processes. Figure~\ref{fig:hww_fig1} shows the $\Delta \phi$
distributions for SM backgrounds, for Higgs masses of 140 and 160
GeV/$c^2$, and for the data. We observe no evidence for a signal over
the SM expectations. We calculate upper limits on the production
cross section times branching ratio, $\sigma_{H} \times {\rm{BR}}(H
\rightarrow WW^\ast$), using a Bayesian procedure. We consider three
components in the data: $H \rightarrow WW^\ast$, SM $WW$, and other SM
processes ($WZ$, $ZZ$, Drell Yan, $W+\rm{jets}$/$\gamma$) labeled as
``other''. The expected number of events in each $\Delta \phi$ bin is
\begin{eqnarray}
\mu = f_{WW}\cdot n_{WW} + f_{other}\cdot n_{other} + \nonumber \\
 f_{HWW} \cdot (\epsilon \cdot {\cal{L}} \cdot \sigma_{H} \cdot BR(H\rightarrow WW^{*})),\nonumber 
\label{eq1}
\end{eqnarray}
where $f_{WW}$, $f_{other}$ and $f_{HWW}$ represent the expected
fraction of the specified categories of events falling in each $\Delta
\phi$ bin, $n_{WW}$ and $ n_{other}$ are the expected numbers of $WW$ and
non-$WW$ background events, and $\epsilon$, $\cal L$ and $\sigma_{H}$
correspond to efficiency, integrated luminosity, and $H$ production
cross section. A posterior density is obtained by multiplying the
Poisson likelihood function with Gaussian prior densities for the
integrated luminosity, background normalizations, and the signal
efficiency:

\begin{figure*}[tbh]
\includegraphics[height=7.6cm,width=0.9\columnwidth]{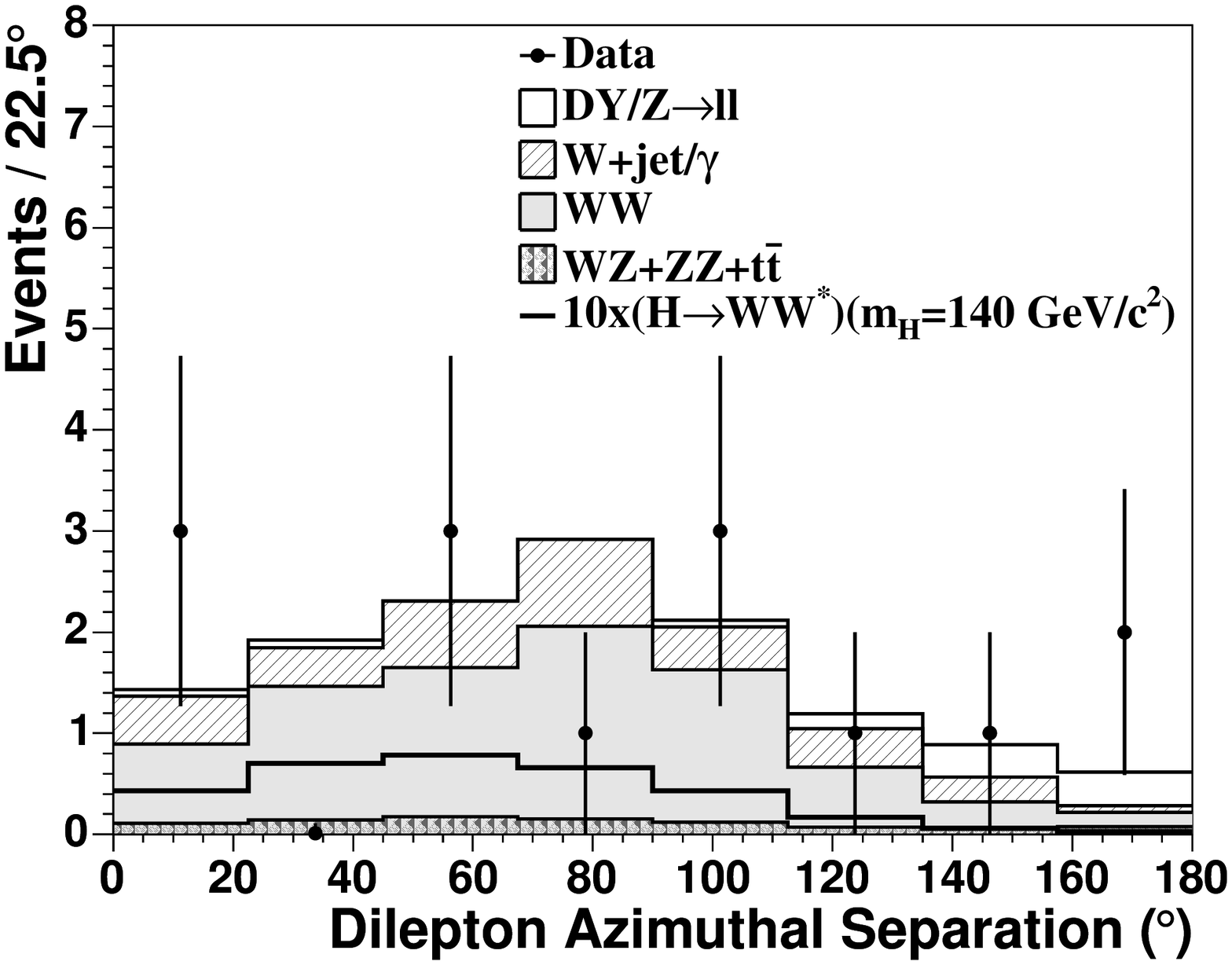}
\includegraphics[height=7.6cm,width=0.9\columnwidth]{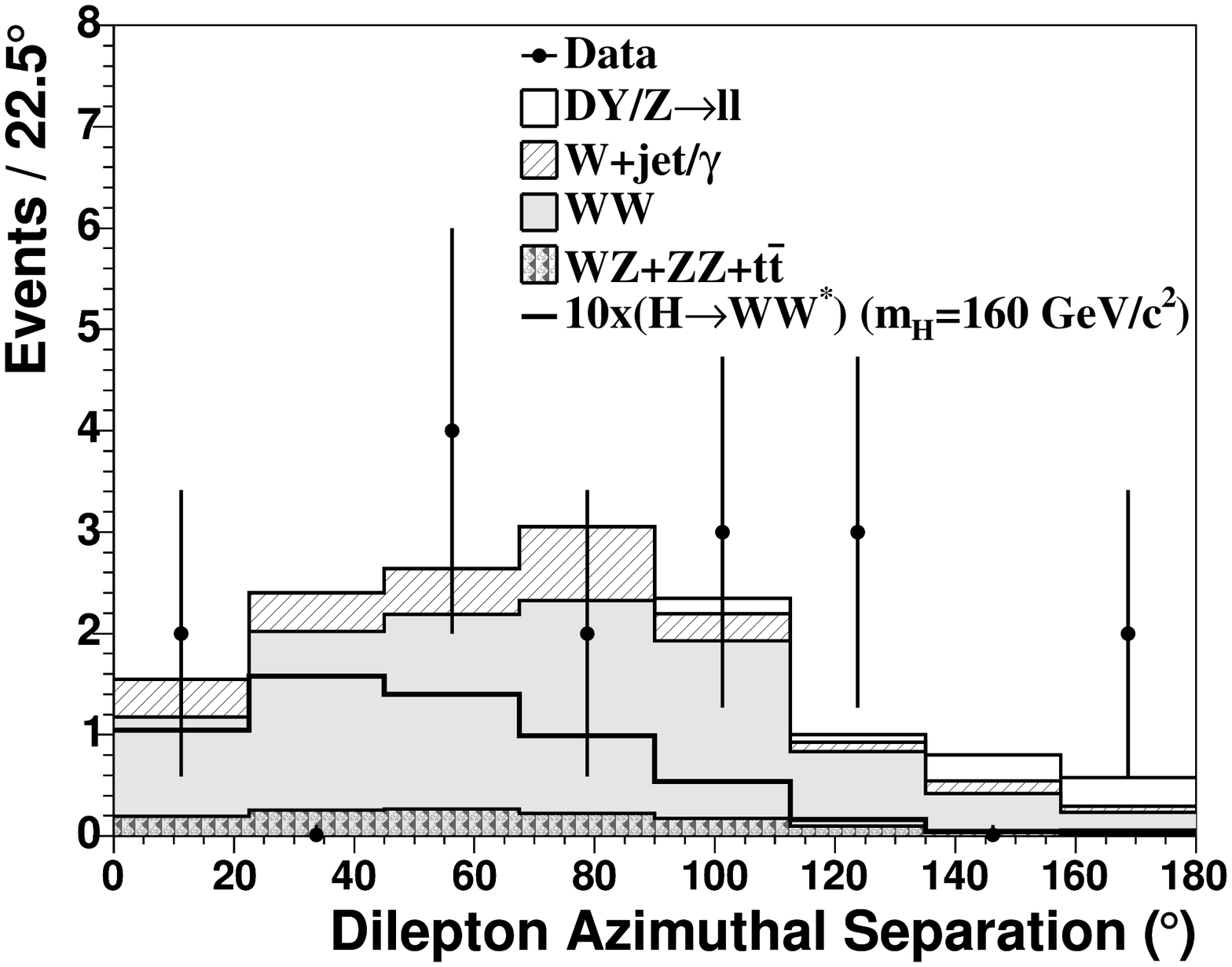}
\caption{\label{fig:hww_fig1} 
Dilepton azimuthal distributions for SM backgrounds, $HWW$ signal and data are shown for 
two Higgs masses: 140 GeV/$c^2$ (left figure) and 160 GeV/$c^2$ (right figure).
 }
\end{figure*}

\begin{eqnarray}
L = \prod_{i=1}^{N_{bins}}{\frac{\mu_{i}^{n_{i}} \cdot e^{-\mu_{i}}}{n_{i}!} \times G(n_{WW},\sigma_{WW}) \times G(n_{other},\sigma_{other})} \nonumber \\ 
\times G(\epsilon,\sigma_{\epsilon}) \times G(\cal L,\sigma_{\cal L})\nonumber 
\label{eq2}
\end{eqnarray}
where $n_{i}$ is the number of events observed in the data, and
$G(n,\sigma_n)$ are Gaussian constraints for parameter $n$ with
uncertainty $\sigma_{n}$. The prior density for
$\sigma \times$ BR($H
\rightarrow WW^\ast$) is assumed uniform. The posterior density is then
integrated over all parameters except for $\sigma \times {\rm{BR}}(H\rightarrow WW^\ast$), for which a 95\% confidence level upper limit
is obtained by calculating the $95^{\rm th}$ percentile of the
resulting distribution.

\begin{figure}[htb]
\includegraphics[height=7.5cm,width=0.95\columnwidth]{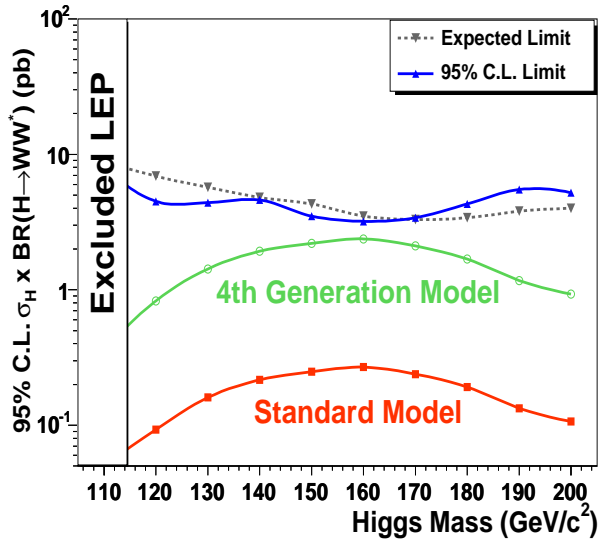}
\caption{\label{fig:vh_fig3}
Summary of the Run II CDF $95\%$ confidence level upper limits on
$\sigma(p\bar{p} \rightarrow H)\times$BR($H\rightarrow WW^{*}$). Shown
for comparison are the standard model prediction, the 4th generation
model prediction~\cite{4gen} and the region excluded by the LEP experiments. The prediction for the 4th generation model assumes that 4th family fermions have a mass $m_{4}$ = 200~GeV/$c^{2}$.}
\end{figure}

The expected and observed upper limits on the cross section times
branching ratio, $\sigma \times {\rm{BR}}(H\rightarrow WW^\ast$), for
different Higgs masses are shown in Table
\ref{table:limits}. The expected limits are calculated using 1000 
simulated experiments, assuming no signal, for each Higgs mass. The
median value of the limits obtained from these experiments is chosen
as the {\it a priori} upper limit.

\begin{table}[h]
\caption{\label{table:limits}
 The expected and observed 95$\%$ C.L. 
limits on $\sigma(p\bar{p}\rightarrow H)\times {\rm{BR}}(H\rightarrow
WW^{*})$. }

\begin{ruledtabular}
\begin{tabular}{lccccc}
$m_{H}$ (GeV/$c^2$) & 120 & 140& 160 & 180 &200 \\ \hline
Expected Limits ($\rm{pb}$)  &  7.1&  4.8&  3.5&  3.4  &  4.0\\ 
Observed Limits ($\rm{pb}$)  &   4.5 & 4.6 &  3.2 &  4.3 &  5.2\\

\end{tabular}
\end{ruledtabular}
\end{table}



In conclusion, observing no signal in the direct search for $H
\rightarrow WW^\ast$, with the subsequent decay of the $W$ bosons to
leptons, we have set mass dependent limits at 95\% C.L. on
$\sigma(p\bar{p}\rightarrow H)\times{\rm{BR}}(H\rightarrow
WW^{*})$. This search is potentially sensitive to other new physics
models such as the example in Figure~\ref{fig:vh_fig3}.


We thank the Fermilab staff and the technical staffs of the participating institutions for their vital contributions. This work was supported by the U.S. Department of Energy and National Science Foundation; the Italian Istituto Nazionale di Fisica Nucleare; the Ministry of Education, Culture, Sports, Science and Technology of Japan; the Natural Sciences and Engineering Research Council of Canada; the National Science Council of the Republic of China; the Swiss National Science Foundation; the A.P. Sloan Foundation; the Bundesministerium f\"ur Bildung und Forschung, Germany; the Korean Science and Engineering Foundation and the Korean Research Foundation; the Particle Physics and Astronomy Research Council and the Royal Society, UK; the Russian Foundation for Basic Research; the Comisi\'on Interministerial de Ciencia y Tecnolog\'{\i}a, Spain; in part by the European Community's Human Potential Programme under contract HPRN-CT-2002-00292; and the Academy of Finland.



\begin{thebibliography}{99}
\bibitem{lumdetail} The integrated luminosities for $ee$, $e\mu$ and $\mu\mu$ are $363 \pm 22$, 
$356 \pm 21$, and $364 \pm 22~{\rm{pb}^{-1}}$ respectively. The differences
for the various channels are due to different detector quality
requirements.
\bibitem{eta} In the CDF coordinate system, $\theta$ and $\phi$ are the 
polar and azimuthal angles, respectively, with respect to the proton
beam direction ($z$ axis). The pseudorapidity $\eta$ is defined as
$-\ln \tan(\theta/2)$.  The transverse momentum of a particle is $p_{T}
= p \sin\theta$.  The analogous quantity using calorimeter energies,
defined as $E_{T} = E \sin\theta$, is called transverse energy.
\bibitem{met}
The missing transverse energy, $\met$, is defined as $|-\sum \Et^i
\,\hat{n}_i|$, where $\hat{n}_i$ is the unit vector in the transverse
plane pointing from the interaction point to the energy deposition in
calorimeter cell $i$.
\bibitem{caliso} The calorimeter isolation $I_{cal}$ is defined as the extra energy deposited 
in the calorimeter cone of radius $\Delta R=\sqrt{{\Delta \phi}^{2}+
{\Delta \eta}^{2}}= 0.4$ around the lepton cluster. The $I_{cal}$ is
required to be less than 10 $\%$ of the lepton $E_{T}$. 
  The track isolation $I_{trk}$ is defined as the sum $p_T$ of all the tracks in a cone of radius 
$\Delta R = 0.4$ around the lepton 
candidate track, but excluding it. The $I_{trk}$ is required to be less than 10$\%$ of the lepton track $p_{T}$.
\bibitem{pdferror} We compared the acceptances obtained using MRST 72 and 75 PDF sets, 
corresponding to $\alpha_{s}$ = 0.1175 and 0.1125. Also we
include the effect of the eigenvector variations around the fit
minima, using CTEQ6M PDF sets.
\end{thebibliography}
\end{document}